\documentclass[times,twocolumn,final]{elsarticle}

\usepackage{arxiv}
\usepackage{tabularx}%
\usepackage{graphicx}%
\usepackage{multirow}%
\usepackage{amsmath,amssymb,amsfonts}%
\usepackage{amsthm}%
\usepackage{mathrsfs}%
\usepackage[title]{appendix}%
\usepackage[table,xcdraw]{xcolor}
\usepackage{textcomp}%
\usepackage{manyfoot}%
\usepackage{booktabs}%
\usepackage{algorithm}%
\usepackage{algorithmicx}%
\usepackage{algpseudocode}%
\usepackage{listings}%
\usepackage{multirow}%
\usepackage{pifont}%
\usepackage{adjustbox}
\usepackage{graphicx}
\usepackage{hyperref}
\usepackage{float}
\usepackage{placeins}
\usepackage{fancyhdr}
\usepackage{threeparttable}
\usepackage{epsfig, xspace, enumitem}
\usepackage{caption, overpic, textpos, pifont, adjustbox}
\definecolor{lightgray}{rgb}{0.9, 0.9, 0.9}

\renewcommand{\headrulewidth}{0pt} 
\begin{document}
\fancypagestyle{firstpagestyle}{
    \fancyhf{} 
    \renewcommand{\headrulewidth}{0pt} 
    \fancyhead{} 
    \fancyfoot{} 
    \fancyhead[CO]{\em \fontsize{9pt}{8pt}\selectfont This article has been accepted to the  International Journal of Computer Assisted Radiology and Surgery.}
}

\fancypagestyle{default}{
    \fancyhf{}
    \fancyhead[R]{\thepage} 
    \renewcommand{\headrulewidth}{0.4pt} 
    \fancyhead[LO]{\em \fontsize{9pt}{8pt}\selectfont This article has been accepted to the  International Journal of Computer Assisted Radiology and Surgery.}
}



\title{UltraSam: A Foundation Model for Ultrasound using Large Open-Access Segmentation Datasets}

\author[1,2]{Adrien \snm{Meyer} \fnref{corresp}}
\fntext[corresp]{Corresponding author: \texttt{ameyer1@unistra.fr}}
\author[1,2]{Aditya \snm{Murali}}
\author[1,2]{Farahdiba \snm{Zarin}}
\author[2,3]{Didier \snm{Mutter}}
\author[1,2]{Nicolas \snm{Padoy}}

\address[1]{University of Strasbourg, CNRS, INSERM, ICube, UMR7357, Strasbourg, France}
\address[2]{IHU-Strasbourg, Institute of Image-Guided Surgery, Strasbourg, France}
\address[3]{Hôpitaux Universitaires de Strasbourg, Strasbourg, France}

\received{XXX}
\finalform{XXX}
\accepted{XXX}
\availableonline{XXX}
\communicated{XXX}

\begin{abstract}
\noindent\textbf{Purpose:} Automated ultrasound (US) image analysis remains a longstanding challenge due to anatomical complexity and the scarcity of annotated data. Although large-scale pretraining has improved data efficiency in many visual domains, its impact in US is limited by a pronounced domain shift from other imaging modalities and high variability across clinical applications, such as chest, ovarian, and endoscopic imaging. To address this, we propose UltraSam, a SAM-style model trained on a heterogeneous collection of publicly available segmentation datasets, originally developed in isolation. UltraSam is trained under the prompt-conditioned segmentation paradigm, which eliminates the need for unified labels and enables generalization to a broad range of downstream tasks.

\noindent\textbf{Methods}: We compile US-43d, a large-scale collection of 43 open-access US datasets comprising over 282,000 images with segmentation masks covering 58 anatomical structures. We explore adaptation and fine-tuning strategies for SAM and systematically evaluate  transferability across downstream tasks, comparing against state-of-the-art pretraining methods. We further propose prompted classification, a new use case where object-specific prompts and image features are jointly decoded to improve classification performance.

\noindent\textbf{Results}: In experiments on three diverse public US datasets, UltraSam outperforms existing SAM variants on prompt-based segmentation and surpasses self-supervised US foundation models on downstream (prompted) classification and instance segmentation tasks.

\noindent\textbf{Conclusion}: UltraSam demonstrates that SAM-style training on diverse, sparsely annotated US data enables effective generalization across tasks. By unlocking the value of fragmented public datasets, our approach lays the foundation for scalable, real-world US representation learning. We release our code and pretrained models at \href{https://github.com/CAMMA-public/UltraSam}{https://github.com/CAMMA-public/UltraSam} and invite the community to further this effort by continuing to contribute high-quality datasets.
\\

\end{abstract}

\maketitle
\thispagestyle{firstpagestyle}

\section{Introduction}\label{sec1}
\begin{figure*}[t!]%
\centering
\includegraphics[width=0.9\textwidth]{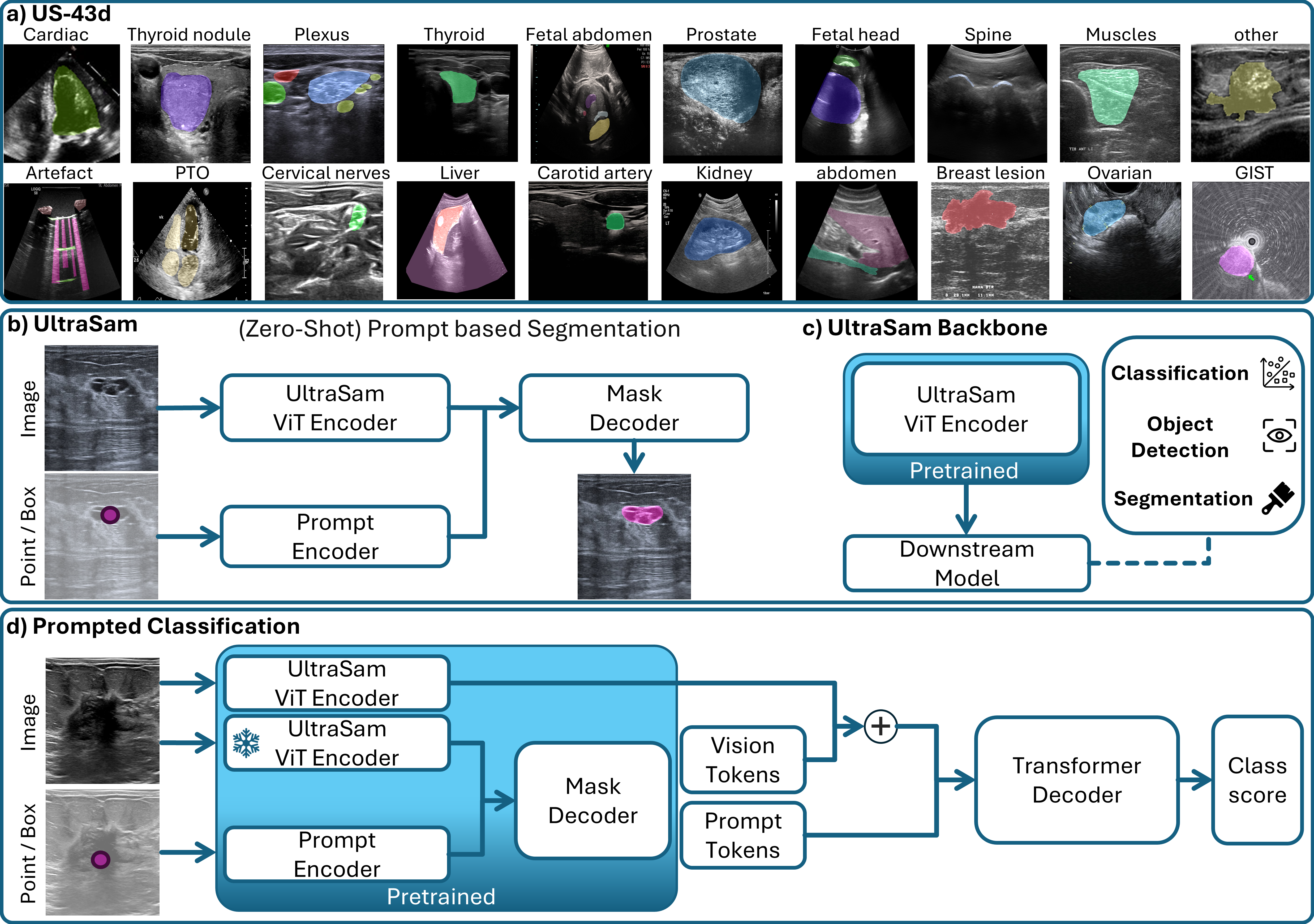}
\caption{UltraSam overview. a) US-43d: a large-scale open US segmentation dataset. b) Fine-tuning SAM on US-43d enables strong zero-shot, prompt-based segmentation. c) UltraSam's pretrained feature extractor provides a robust foundation for downstream tasks. d) We propose prompted classification to enhance structure classification using a user-specified prompt.}\label{figOverview}
\end{figure*}

\begin{figure*}[t]%
\centering
\includegraphics[width=0.9\textwidth]{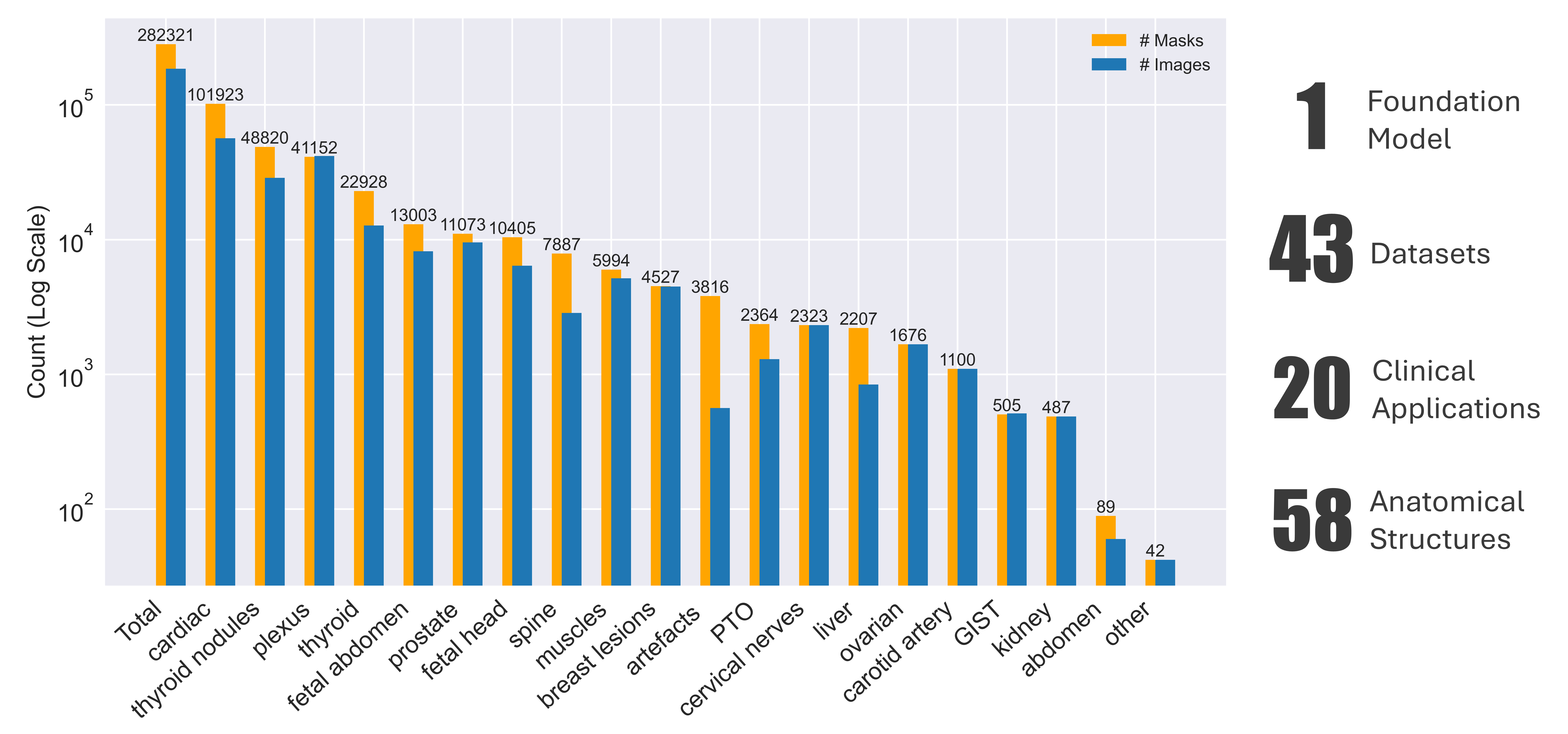}
\caption{Overview of US-43d, grouped by clinical applications. PTO refers to patent foramen ovale, and GIST refers to gastrointestinal stromal tumor.}\label{fig:dataset_snapshot}
\end{figure*}
Ultrasound (US) has become indispensable in modern medicine as a real-time, safe, and cost-effective imaging technique. It plays a crucial role in dynamic assessments, such as fetal monitoring, and its portability makes it accessible even in low-resource settings, significantly enhancing the reach of diagnostic care.
However, interpreting US images is often challenging due to factors like noise and variability, and therefore requires highly-skilled practitioners.
Assistive computer vision solutions have emerged as a general approach to ease US image analysis, with successful applications ranging from anatomical landmark identification, to tissue characterization from digital biopsy, to needle tracking during interventional procedures~\cite{brachial_plexus,mmotu}.
While promising, most existing solutions are task-specific and evaluated on small benchmark datasets; scaling these methods to diverse clinical settings is still an open problem.

In this vein, both the general and medical computer vision communities have begun to shift towards the development and application of foundation models trained on diverse data, which can, in concept, ensure strong generalization capabilities both when used out-of-the-box and when serving as model initializations.
However, general-purpose or medical foundation models tend to be ineffective on US images due to substantial domain shift, while the dataset scale required to train US-specific foundation models can only be achieved by combining highly heterogeneous images originating from numerous examination areas (e.g. chest, ovarian, endoscopic).
A few works have tackled the latter, proposing frameworks to train US-specific foundation models through self-supervised learning (SSL), a training paradigm that leverages unlabeled data to learn useful representations~\cite{jiao2024usfm,kang2023deblurring}, and by leveraging labels when available~\cite{SAMUS,chen2024multi}; still, ensuring effective generalization to diverse organ types remains a significant challenge.

Our key insight is that Segment-Anything Model (SAM)-style training can produce improved foundation models by better leveraging diverse ultrasound data.
Because SAMs predict segmentation masks based on instance-specific prompts, such as points or bounding boxes, rather than pre-defined classes, they are naturally suited to handle diverse datasets with non-overlapping classes and sparsely annotated instances; as a result, they learn a rich object-centric representation space that could greatly aid various downstream tasks.
A few works have focused on building a US-specific SAM: SAMUS~\cite{SAMUS} compiles US30K from seven public datasets, and trains an adapter on top of SAM; SonoSAM~\cite{sonosam} is a closed-source finetune of SAM; and BUSSAM~\cite{BUSSAM} adapts SAM for breast lesion segmentation.
Yet, of these works, only SAMUS is an open-source general-purpose SAM adaptation for US, and it is only trained on a relatively small-scale dataset (30K masks).
Moreover, all of these works limit their evaluation to segmentation; as a result, it is difficult to gauge their foundational capabilities.

To tackle these shortcomings, we begin by addressing dataset scale, compiling US-43d, a collection of 282,321 image-segmentation mask pairs from 43 public datasets.
We then train UltraSam by fully-finetuning SAM on US-43d, showing through a series of evaluations and comparisons against existing Medical SAMs (e.g. MedSAM~\cite{medSam}, Medical SAM Adapter (Med-SA)~\cite{MSA}, SAMUS~\cite{SAMUS}) that
UltraSam is a far more robust and powerful interactive segmentation model for US, even on completely unseen organs.
Then, we benchmark downstream instance segmentation and classification performance, finetuning each of the SAMs as well as other US foundation models~\cite{MSA}.
Finally, we introduce prompted classification, a natural extension of the SAM architecture that improves downstream classification by explicitly leveraging user-specified point or box prompts; this task is particularly relevant in medical image analysis, where `detect-then-classify' tasks - e.g. a digital biopsy to identify a lesion then characterize it - are commonplace.

In summary, our contributions are as follows:
\begin{enumerate}
  \item We compile and release the largest publicly available collection of ultrasound segmentation data, US-43d, consisting of 43 datasets and 282,321 pairs of images and masks, covering 20 different clinical applications.
  \item We introduce UltraSam, a state-of-the-art SAM for ultrasound images.
  \item We demonstrate the superior foundational capabilities of UltraSam compared to existing medical SAMs and US foundation models through downstream instance segmentation and image classification experiments.
  \item We introduce a novel use case for SAMs, prompted classification, and show that it outperforms traditional downstream classification.
\end{enumerate}

\section{Methods}

\subsection{Dataset}\label{subsec2}
US imaging presents a substantial domain gap compared to other medical imaging modalities; building an US-specific foundation model therefore requires a specialized large-scale dataset. To build such a dataset, we crawl a multitude of platforms for human medical US with instance annotations and open-access availability: Papers with Code, Google Dataset Search, GitHub, Google Scholar, Kaggle, ResearchGate, Mendeley dataset, Zenodo and Data in Brief.
Through this process, we arrive at US-43d (see Fig.\ref{figOverview}.a), a collection of 43 datasets covering 20 different clinical applications, containing 282,321 annotated segmentation masks from both 2D and 3D scans. US-43d captures organs and lesion of various shapes, sizes, and textures across clinical applications such as cardiac, fetal head, thyroid, and breast lesions, as illustrated in Fig.\ref{fig:dataset_snapshot}, providing a comprehensive view of the medical ultrasound landscape. Table~\ref{appendix:detailled_datasets} provides detailed information on the US-43d dataset, including dataset names, access links, and the number of images and segmentation masks available in each.

For testing, we select three diverse datasets from US-43d: BUS-BRA~\cite{BUSBRA} (breast lesions, 1875 images), MMOTU2D~\cite{mmotu} (ovarian lesions, 1469 images), and GIST514-DB~\cite{GIST514} (gastrointestinal stromal tumors, 514 images). GIST514-DB is included as an outlier in our selection, as it is the only dataset in US-43d with radial acquisition. We evaluate on the official test split of each dataset.
Together they include linear and radial probes, endoscopic and non-endoscopic US, and span multiple clinical applications, anatomical regions, lesion types, and imaging techniques, enabling exhaustive evaluation of UltraSam's generalizability.
We reserve 5\% of each training dataset for validation and use the remaining 95\% for training. We preprocess images by removing label-background overlaps (common in 3D US), cropping backgrounds occupying more than 50\% of image pixels, and using sagittal views for 3D images.

\begin{table}
\scriptsize
\begin{center}
\caption{Overview and links of the US-43d ultrasound datasets.}\label{appendix:detailled_datasets}

\begin{tabular}{@{\extracolsep{\fill}}cccc}
\toprule
Dataset (Link) &  Clinical Applications & \# Images & \# Masks \\
\midrule
\href{https://github.com/Regional-US/brachial_plexus}{Brachial Plexus} & plexus & $40788$ & $36736$ \\

\href{https://github.com/echonet/dynamic}{EchoNet-Dynamic} & cardiac & $20048$ & $20048$ \\

\href{https://humanheart-project.creatis.insa-lyon.fr/database/#collection/6373703d73e9f0047faa1bc8g}{CAMUS} & cardiac & $19232$ & $58570$ \\

\href{https://stanfordaimi.azurewebsites.net/datasets/a72f2b02-7b53-4c5d-963c-d7253220bfd5}{Thyroid US CineClip} & thyroid nodules & 17412 & 17412 \\

\href{https://echonet.github.io/pediatric}{Echonet Pediatric} & cardiac & 15449 & 15449 \\

\href{https://www.cs.cit.tum.de/camp/publications/segthy-dataset/}{Segthy-Dataset} & thyroid & 12737 & 22928 \\

\href{https://acouslic-ai.grand-challenge.org/overview-and-goals/}{ACOUSLIC} & fetal abdomen & 6620 & 6620 \\

\href{https://www.kaggle.com/c/ultrasound-nerve-segmentation/data}{US Nerve Segmentation} & cervical nerves & 5635 & 2323 \\

\href{https://muregpro.github.io/data.html}{regPro} & prostate & 4706 & 6492 \\

\href{https://data.mendeley.com/datasets/3jykz7wz8d/1}{STMUS NDA} & muscles & 4355 & 4368 \\

\href{https://github.com/vahidashkani/Fast-U-Net}{FH-PS-AOP} & fetal head & 4000 & 7999 \\
\href{https://github.com/openmedlab/Awesome-Medical-Dataset/blob/main/resources/TN-SCUI2020.md}{TNSCUI} & thyroid nodules & 3644 & 3659 \\

\href{https://github.com/openmedlab/Awesome-Medical-Dataset/blob/main/resources/TN3K.md}{TG3K} & thyroid nodules & 3585 & 23283 \\

\href{https://github.com/openmedlab/Awesome-Medical-Dataset/blob/main/resources/TN3K.md}{TN3K} & thyroid nodules & 3493 & 3821 \\

\href{https://zenodo.org/records/10475293}{MUP \& MicroSeg} & prostate & 2910 & 2650 \\

\href{https://www.ncbi.nlm.nih.gov/pmc/articles/PMC7654705/}{ASUS} & spine & 2864 & 7887 \\

\href{https://www.kaggle.com/datasets/xiaoweixumedicalai/cardiacudc-dataset}{CardiacUDC} & cardiac & 1961 & 7251 \\

\href{https://github.com/wgomezf/BUS-BRA}{BUS-BRA} & breast lesions & 1875 & 1875 \\
\href{https://data.mendeley.com/datasets/4gcpm9dsc3/1}{FASS} & fetal abdomen & 1588 & 6383 \\

\href{https://github.com/cv516Buaa/MMOTU_DS2Net}{MMOTU 2d} & ovarian & 1489 & 1489 \\

\href{https://github.com/vahidashkani/Fast-U-Net}{Fast-U-Net} & fetal head & 1411 & 1407 \\

\href{https://www.kaggle.com/datasets/xiaoweixumedicalai/echocp}{EchoCP} & Patent Foramen Ovale & 1300 & 2364 \\

\href{https://data.mendeley.com/datasets/d4xt63mgjm/1}{Common Carotid Artery} &  carotid artery & 1100 & 1100 \\

\href{https://hc18.grand-challenge.org/}{HC18} & fetal head & 999 & 999 \\

\href{https://ubpd.worldwidetracing.com:9443/}{UBPD} & plexus & 939 & 4416 \\

\href{https://kalisteo.cea.fr/index.php/fallmud/#}{FALLMUD} & muscles & 810 & 1626 \\

\href{https://data.mendeley.com/datasets/3ksd7w7jkx/1}{BUS-UC} & breast lesions & 810 & 791 \\

\href{https://zenodo.org/records/7272660}{AUL} & liver & 735 & 2102 \\

\href{https://scholar.cu.edu.eg/?q=afahmy/pages/dataset}{Breast} & breast lesions & 690 & 690 \\

\href{https://github.com/openmedlab/Awesome-Medical-Dataset/blob/main/resources/TN3K.md}{DDTI} & thyroid nodules & 637 & 645 \\

\href{https://archive.researchdata.leeds.ac.uk/1263/}{LUSS phantom} & artefacts & 564 & 3816 \\

\href{https://github.com/howardchina/query2}{GIST514-DB} & GIST & 514 & 505 \\

\href{https://github.com/rsingla92/kidneyUS}{KidneyUS} & kidney & 487 & 487 \\

\href{https://data.mendeley.com/datasets/7fvgj4jsp7/1}{BUS-UCLM} & breast lesions & 264 & 281 \\

\href{https://www.cancerimagingarchive.net/collection/breast-lesions-usg/}{BrEaST} & breast lesions & 252 & 266 \\

\href{https://qamebi.com/breast-ultrasound-images-database/}{BUID} & breast lesions & 232 & 236 \\

\href{https://www.ncbi.nlm.nih.gov/pmc/articles/PMC8205136/}{S1} & breast lesions & 201 & 204 \\

\href{https://github.com/cv516Buaa/MMOTU_DS2Net}{MMOTU 3d} & ovarian & 187 & 187 \\

\href{http://www2.docm.mmu.ac.uk/STAFF/M.Yap/dataset.php}{BUS} & breast lesions & 164 & 164 \\

\href{https://www.researchgate.net/publication/329586355_100_2D_US_Images_and_Tumor_Segmentation_Masks}{105US} & liver & 105 & 105 \\

\href{https://www.kaggle.com/datasets/ignaciorlando/ussimandsegm}{AbdomenUS} & abdomen & 60 & 89 \\

\href{https://github.com/xbhlk/STU-Hospital}{STU-Hospital} & other & 42 & 42 \\

\bottomrule
\end{tabular}
\end{center}
\end{table}

\subsection{UltraSam}

We adopt the architecture of SAM~\cite{sam}, depicted in Fig.~\ref{figOverview}.b, which utilizes a 12 layers Vision Transformer encoder to extract image features as tokens. A prompt encoder transforms prompts, such as points or boxes, into object query tokens. These tokens interact with the image feature tokens through a 2 layers transformer decoder, enabling reasoning and interaction between prompts and vision tokens. A mask head predicts multiple mask outputs using an MLP, each with a corresponding predicted Intersection over Union (IoU) score, allowing selection of the best predicted mask. In an additional pass through the decoder, the mask logits from the previous iteration are encoded and added element-wise to the image embedding, refining the mask prediction.

\subsection{Prompted Classification.} 
Building on SAM’s paradigm of instance-specific prompts for segmentation, we extend this concept to prompt-based classification (Fig.~\ref{figOverview}.d), where a point or box prompt enables instance-level classification within an image. This approach leverages UltraSam’s object-centric ViT architecture, designed for segmentation, and adapts it for classification while preserving its instance-awareness. A frozen encoder (\(E_o\)), initialized with UltraSam’s weights, extracts object-centric embeddings. A trainable encoder (\(E_g\)) produces semantic-rich image-level embeddings (\(z_g\)) for classification. The object-centric tokens from \(E_o\) interact with prompt tokens through the mask decoder, generating fine-grained instance representations. Vision tokens from \(E_o\) are added with global tokens from \(E_g\), and the resulting features, along with prompt tokens, are passed to a transformer decoder containing an additional classification token. This setup allows interaction across all tokens, enabling UltraSam to integrate instance- and image-level cues. The final classification token is concatenated with \(z_g\) to predict the class score.

During training for both interactive segmentation and prompted classification, we simulate user prompts by randomly sampling either a point or a box with equal probability for each instance. The point is selected randomly within the instance mask, while the box is a noised version of the ground truth (GT) box annotation. To generate this noise, the two box corners are randomly displaced by up to 5 percent of the box's width and height. For evaluation, we follow SAM's approach and report results using either the center point or the GT box as prompt. We also evaluate UltraSam for downstream tasks (see Fig.~\ref{figOverview}.c), leveraging its feature extractor as a pretrained backbone for our models.

\subsection{Implementation details} 
We initialize UltraSam using the pretrained SAM ViT-b model then finetune on four H100 GPUs for 30k iterations with a batch size of eight images per GPU (16 hours total training time). Images are resized then padded to 1024x1024, maintaining aspect ratio. Our code is based on MMDetection v3.3. We use the AdamW optimizer, with an initial learning rate of $1 \times 10^{-4}$ and a warm-up period of 500 iterations; we reduce the learning rate by a factor of 10 at 20k and 28k iterations. 
Following SAM~\cite{sam}, we use a combination of focal and dice loss for segmentation, and L1 loss for IoU prediction (20:1:1).

\begin{table*}[h]

\caption{(1) Fine-tuning SAM using adapters and end-to-end methods. (2) Zero-shot evaluation of interactive segmentation models (mAP, \%). Med-SA: Medical SAM Adapter~\cite{MSA}. "-" indicates unsupported prompts. A: adaptation, ZS: zero-shot, FT: end-to-end Finetuning.}\label{tab:prompt_seg}
\begin{tabularx}{\textwidth}{@{\extracolsep{\fill}}ccccccccc}
    \toprule
    \multirow{3}{*}{Prompt} & \multirow{3}{*}{Method}& \multirow{3}{*}{Category} & \multicolumn{2}{c}{BUS-BRA} & \multicolumn{2}{c}{MMOTU2D} & \multicolumn{2}{c}{GIST514-DB} \\
    \cmidrule(lr){4-5} \cmidrule(lr){6-7}
    \cmidrule(lr){8-9}
    & & & mAP & mAP@50 & mAP & mAP@50 & mAP & mAP@50 \\

    \bottomrule
    \multicolumn{9}{c}{\cellcolor{blue!11}(1) finetuning SAM with adapters and end-to-end methods} \\
    
    \multirow{5}{*}{Point}

    & LoRa & A
    & $58.5$ & $95.0$
    & $44.4$ & $76.1$
    & $36.9$ & $70.3$\\

    & Med-SA & A
    & $\underline{64.4}$ & $\textbf{96.7}$
    & $\underline{55.9}$ & $\underline{85.9}$
    & $\underline{52.0}$ & $\underline{90.7}$\\

    & SAMUS & A
    & $51.8$ & $91.4$
    & $40.2$ & $70.7$
    & $31.2$ & $73.2$\\
    

    & UltraSam & FT
    & $\textbf{67.1}$ & $\textbf{96.7}$
    & $\textbf{58.2}$ & $\textbf{86.7}$  
    & $\textbf{55.5}$ & $\textbf{90.8}$\\

    \midrule
    \multirow{5}{*}{Box}
  
    & LoRa & A
    & $72.9$ & $\textbf{99.0}$
    & $75.3$ & $\underline{99.0}$
    & $66.7$ & $\underline{99.0}$\\
    
    & Med-SA & A
    & $\underline{78.0}$ & $\textbf{99.0}$
    & $\textbf{79.5}$ & $\underline{99.0}$
    & $\underline{70.0}$ & $97.8$\\

    & SAMUS & A
    & $-$ & $-$
    & $-$ & $-$
    & $-$ & $-$\\

    & UltraSam & FT
    & $\textbf{79.1}$ & $\textbf{99.0}$
    & $\textbf{79.5}$ & $\textbf{100}$
    & $\textbf{73.0}$ & $\textbf{100}$\\

\bottomrule
\multicolumn{9}{c}{\cellcolor{blue!11}(2) zero-shot evaluation} \\

    \multirow{3}{*}{Point}

    & SAM & ZS
    & $14.2$ & $27.2$
    & $2.1$ & $4.7$
    & $6.2$ & $12.2$\\

    & MedSAM & ZS
    & $-$ & $-$
    & $-$ & $-$
    & $-$ & $-$\\

    & UltraSam* & ZS
    & $\textbf{58.3}$ & $\textbf{92.7}$
    & $\textbf{44.4}$ & $\textbf{70.6}$
    & $\textbf{9.3}$ & $\textbf{17.4}$\\

    \midrule
    \multirow{3}{*}{Box}

    & SAM & ZS
    & $\underline{68.1}$ & $\textbf{100}$
    & $34.5$ & $58.8$
    & $\underline{57.2}$ & $\underline{92.2}$\\

    & MedSAM & ZS
    & $59.1$ & $98.9$
    & $\underline{48.8}$ & $\underline{95.4}$
    & $30.4$ & $81.0$\\

    & UltraSam* & ZS
    & $\textbf{76.5}$ & $\textbf{100}$
    & $\textbf{79.6}$ & $\textbf{100}$
    & $\textbf{69.4}$ & $\textbf{98.9}$\\
\bottomrule
\end{tabularx}

\end{table*}

\begin{figure*}[t!]%
\centering
\includegraphics[width=0.48\linewidth]{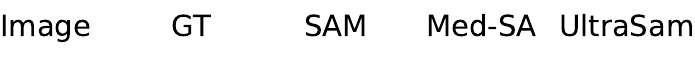}
\hspace{0.2cm}
\includegraphics[width=0.48\linewidth]{text_row.pdf}
\includegraphics[width=0.09\linewidth]{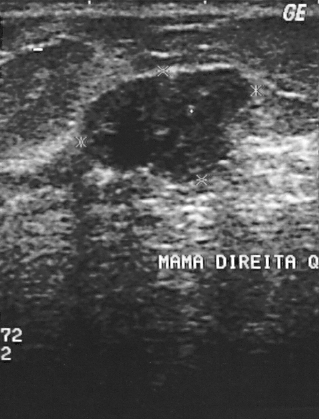}
\includegraphics[width=0.09\linewidth, trim=0 0 319 0, clip]{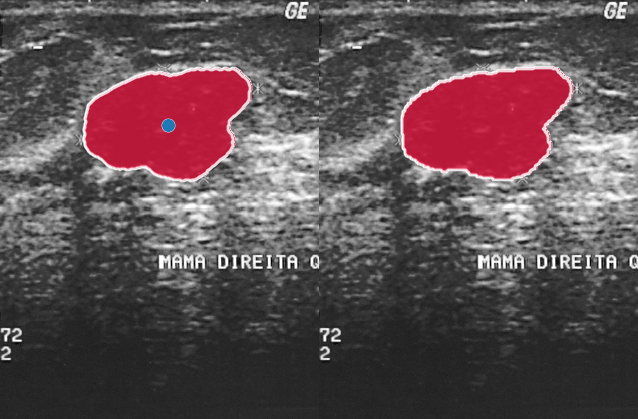}
\includegraphics[width=0.09\linewidth, trim=319 0 0 0, clip]{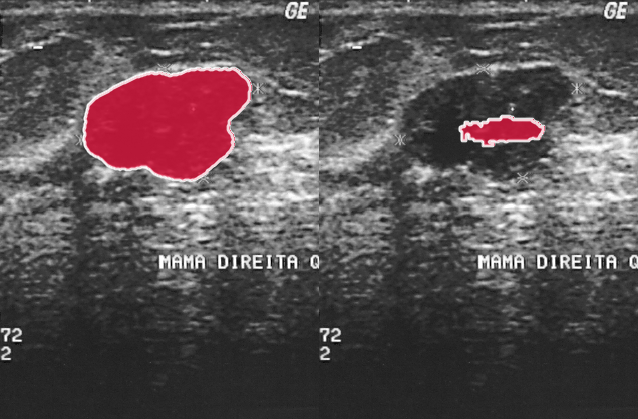}
\includegraphics[width=0.09\linewidth, trim=319 0 0 0, clip]{MEDSA_test.BUSBRA__00373.png}
\includegraphics[width=0.09\linewidth, trim=319 0 0 0, clip]{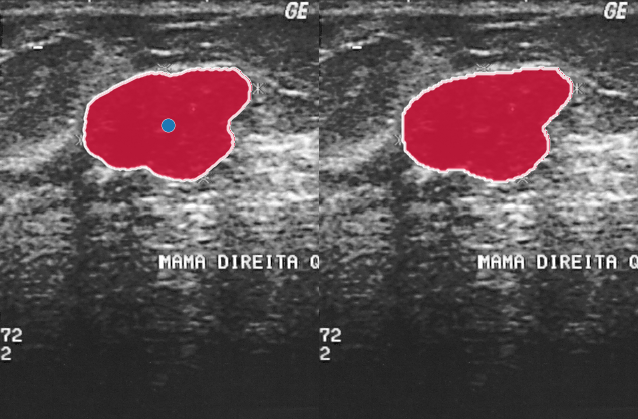}
\hspace{0.15cm}
\includegraphics[width=0.09\linewidth]{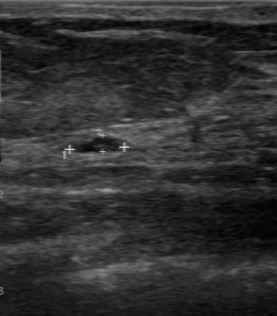}
\includegraphics[width=0.09\linewidth, trim=0 0 277 0, clip]{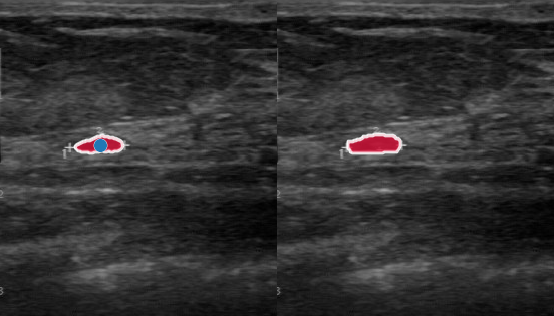}
\includegraphics[width=0.09\linewidth, trim=277 0 0 0, clip]{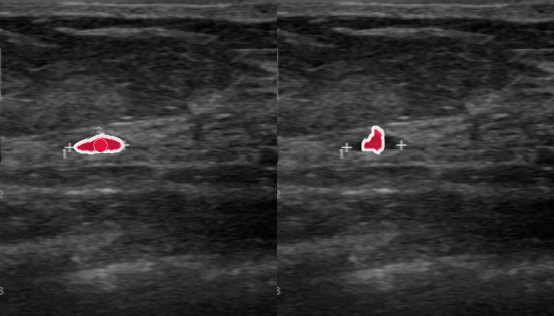}
\includegraphics[width=0.09\linewidth, trim=277 0 0 0, clip]{MEDSA_test.BUSBRA2__01453.png}
\includegraphics[width=0.09\linewidth, trim=277 0 0 0, clip]{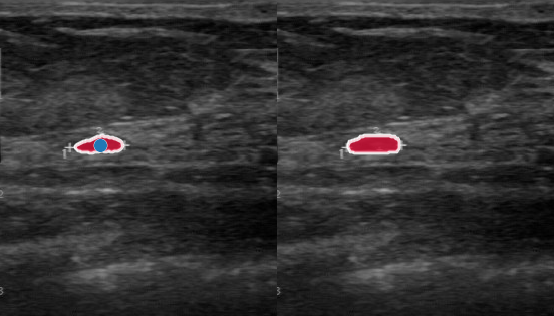}
\includegraphics[width=0.09\linewidth]{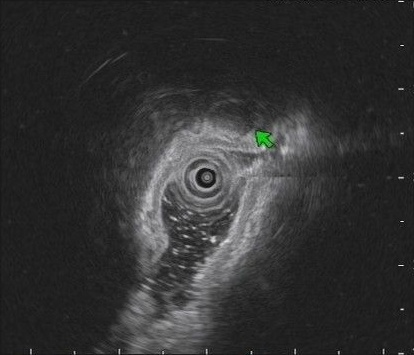}
\includegraphics[width=0.09\linewidth, trim=0 0 414.5 0, clip]{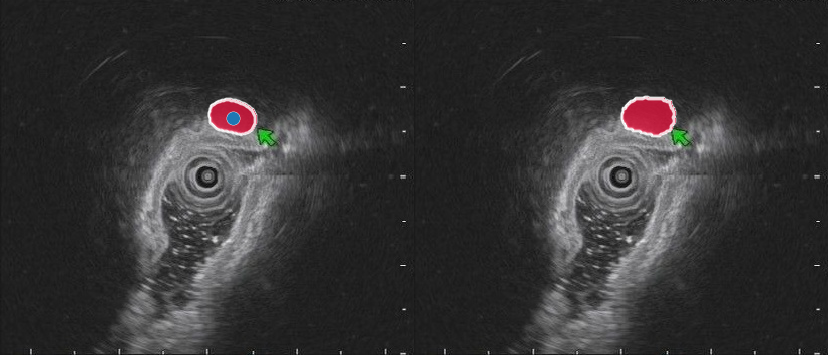}
\includegraphics[width=0.09\linewidth, trim=414.5 0 0 0, clip]{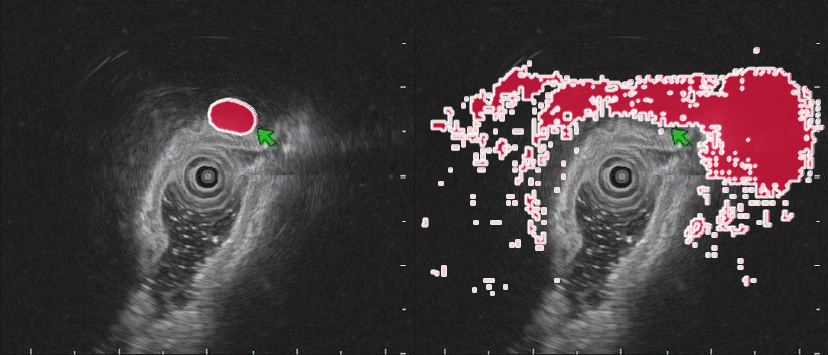}
\includegraphics[width=0.09\linewidth, trim=414.5 0 0 0, clip]{MEDSA_test.GIST514-DB__00006.png}
\includegraphics[width=0.09\linewidth, trim=414.5 0 0 0, clip]{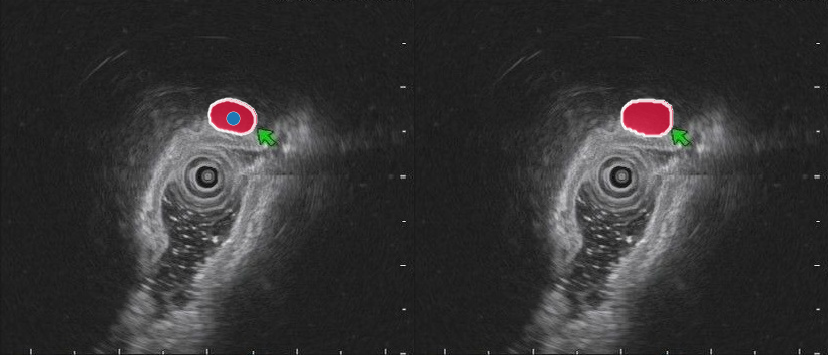}
\hspace{0.15cm}
\includegraphics[width=0.09\linewidth]{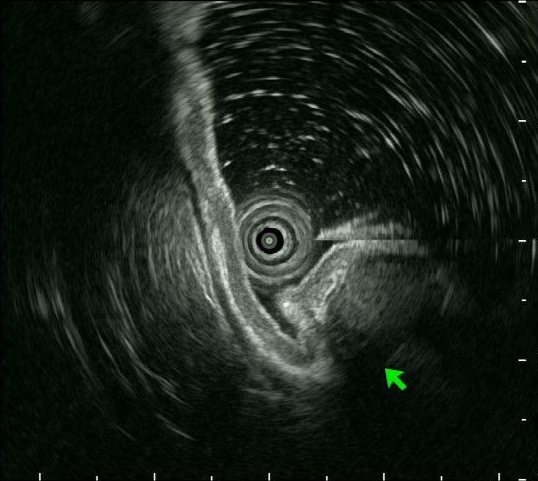}
\includegraphics[width=0.09\linewidth, trim=0 0 538 0, clip]{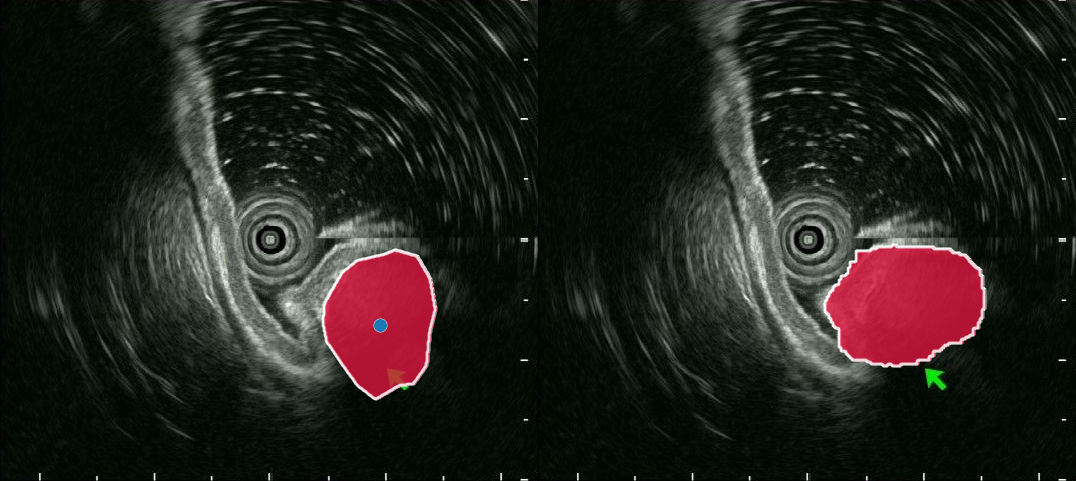}
\includegraphics[width=0.09\linewidth, trim=538 0 0 0, clip]{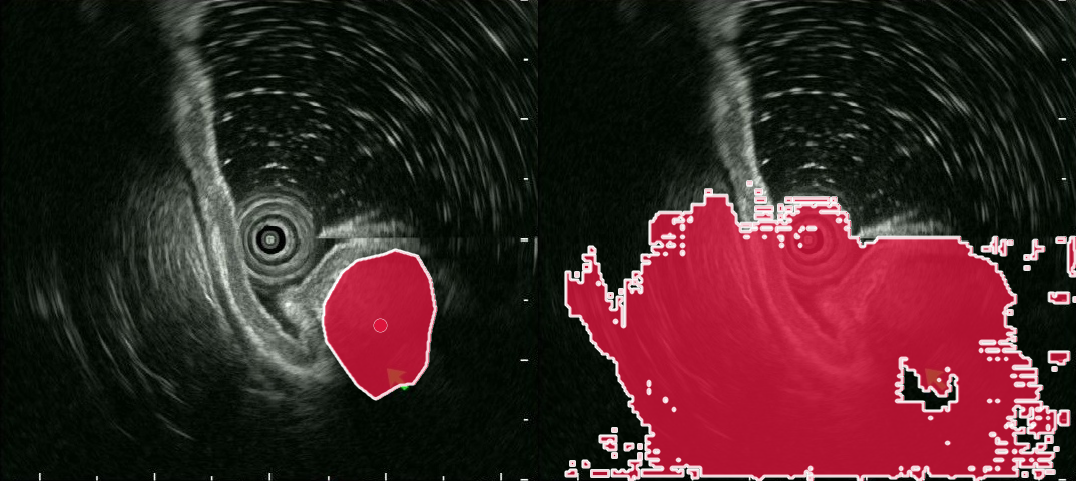}
\includegraphics[width=0.09\linewidth, trim=538 0 0 0, clip]{MEDSA_test.GIST514-DB2.png}
\includegraphics[width=0.09\linewidth, trim=538 0 0 0, clip]{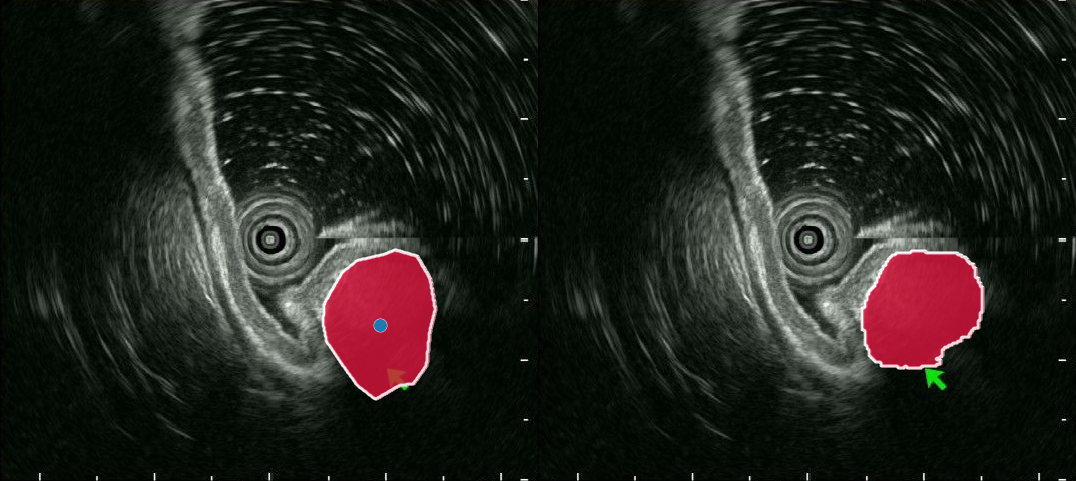}
\includegraphics[width=0.09\linewidth]{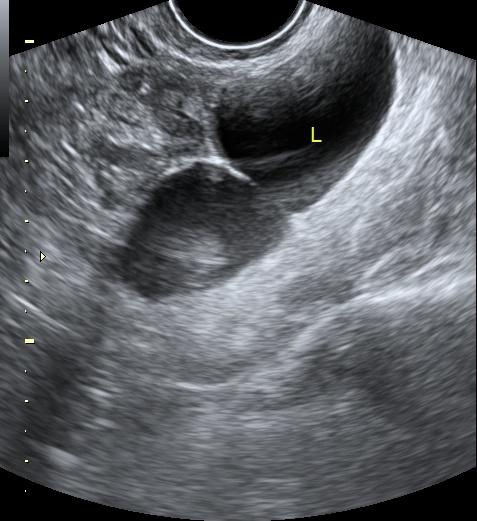}
\includegraphics[width=0.09\linewidth, trim=0 0 477 0, clip]{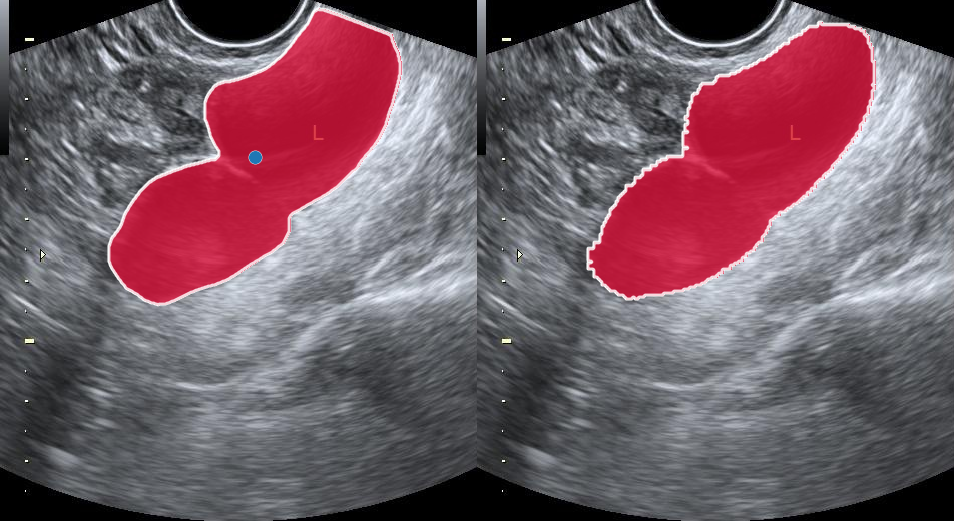}
\includegraphics[width=0.09\linewidth, trim=477 0 0 0, clip]{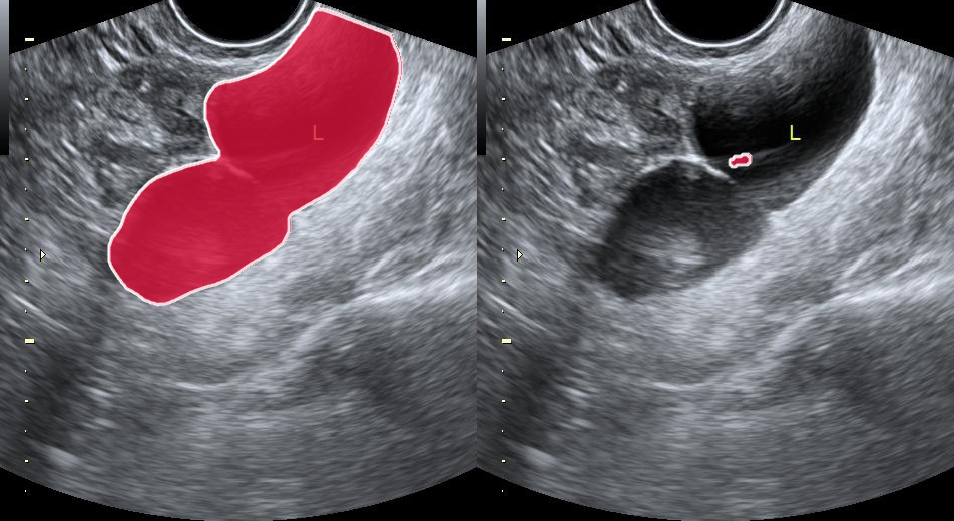}
\includegraphics[width=0.09\linewidth, trim=477 0 0 0, clip]{MEDSA_test.MMOTU_2d__00016.png}
\includegraphics[width=0.09\linewidth, trim=477 0 0 0, clip]{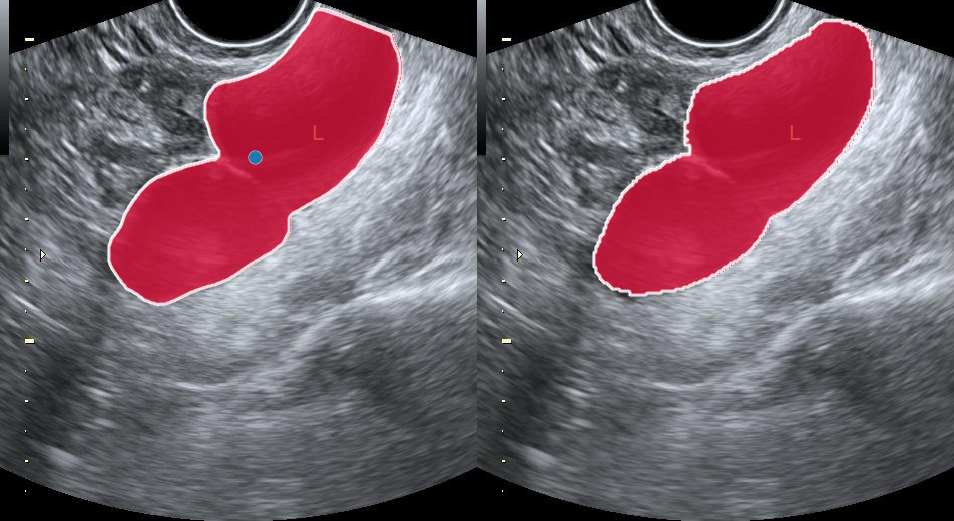}
\hspace{0.15cm}
\includegraphics[width=0.09\linewidth]{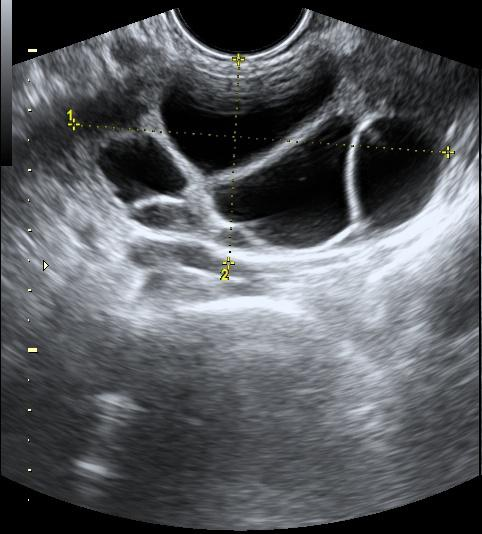}
\includegraphics[width=0.09\linewidth, trim=0 0 482 0, clip]{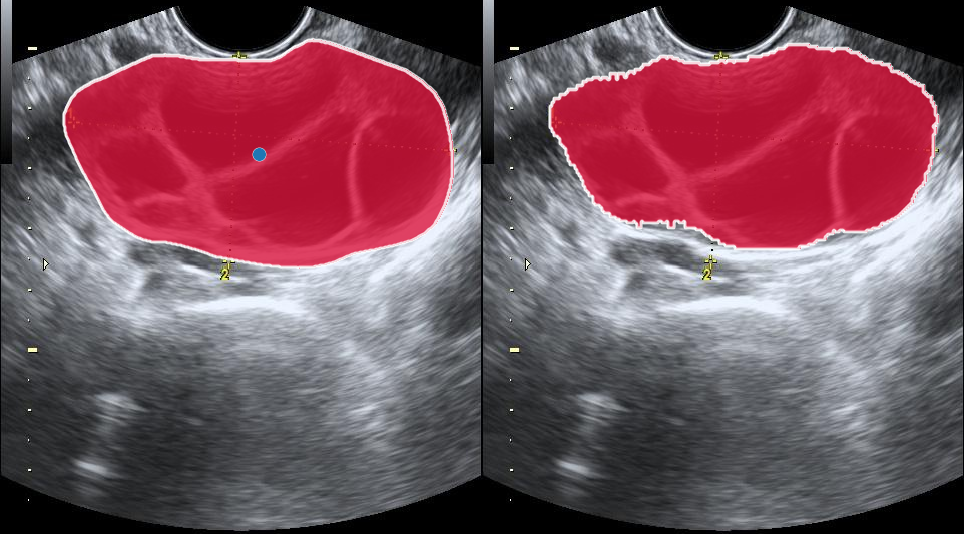}
\includegraphics[width=0.09\linewidth, trim=482 0 0 0, clip]{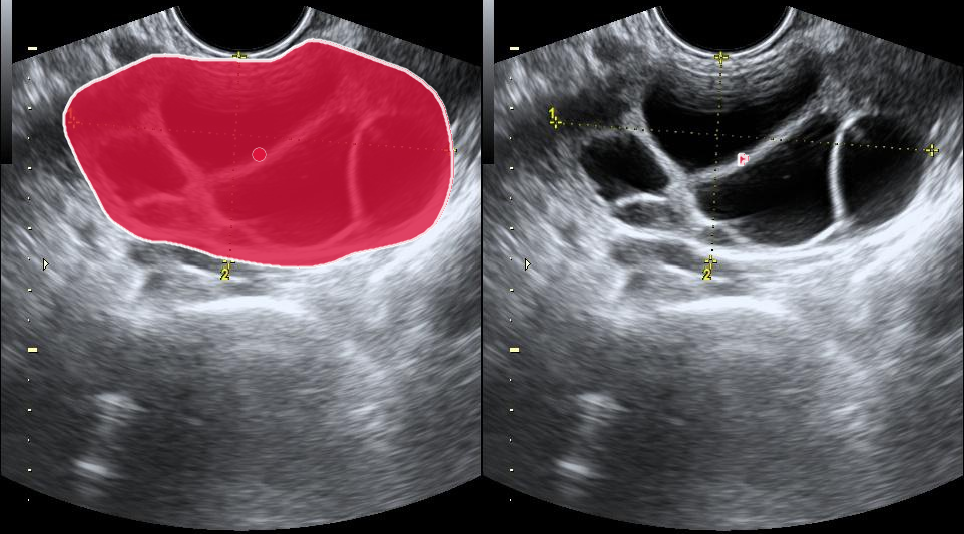}
\includegraphics[width=0.09\linewidth, trim=482 0 0 0, clip]{MEDSA_test.MMOTU_2d2.png}
\includegraphics[width=0.09\linewidth, trim=482 0 0 0, clip]{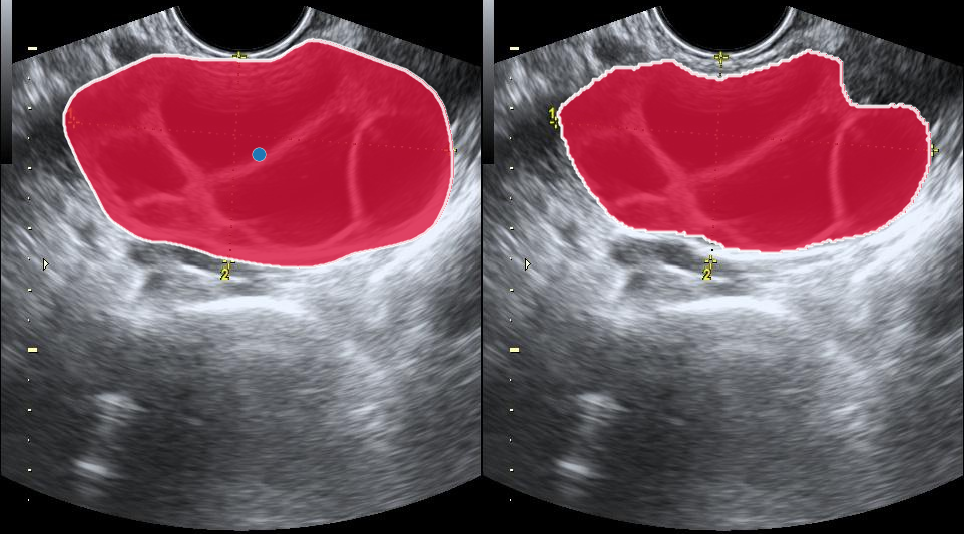}
\caption{Qualitative results for interactive segmentation with a single point-prompt.}\label{quali}
\end{figure*}

\subsection{Evaluation}

\subsubsection{Prompted Evaluation} We evaluate UltraSam for prompt-based segmentation and classification using GT center points or boxes as prompts, comparing it to SAM variants. To assess zero-shot performance, we train dataset-specific variants, denoted as UltraSam*, excluding any datasets containing the target organs\footnote{During training, UltraSam*-BUSBRA excludes all breast US images, UltraSam*-MMOTU2D excludes all ovarian US, and UltraSam*-GIST514DB excludes all gastrointestinal US.}.

\subsubsection{Downstream Task Evaluation} In addition to enabling prompt-based segmentation, UltraSam's feature extractor serves as a powerful pretrained ViT for US. To evaluate its capabilities, we test it on two downstream tasks: instance segmentation and image classification.
For instance segmentation, we use the state-of-the-art Mask2Former~\cite{mask2former}, replacing its Resnet backbone with UltraSam's ViT. For image classification, we build a simple classifier using UltraSam's ViT as the model backbone.
We average the output tokens and apply a linear classifier to predict the label. We compare its performance against SAM and MedSAM’s ViT backbones, the ImageNet-pretrained ResNet-50, self-supervised ultrasound foundation models~\cite{jiao2024usfm,kang2023deblurring}, and dinov2~\cite{dinov2} pretrained on US-43d. The Mask2Former decoder is randomly initialized. We use the default hyperparameters of~\cite{mmdet} and train the models for 8k iterations with a batch size of 8 on a single A100 GPU. All downstream experiments are fine-tuned end-to-end, except for DINOv2, where freezing the backbone was found to significantly improve performance.

\begin{table*}[t]

\caption{Instance segmentation and object detection (mAP, \%) using Mask2Former~\cite{mask2former}, and image classification (Precision (prec), Recall, F1, \%). ResNet-50 initialized with ImageNet weights. US pretrained models are in \colorbox{lightgray}{gray}.}\label{tab:downstream_seg}

\begin{tabularx}{\textwidth}{@{\extracolsep{\fill}}c c cc cc ccc}
\toprule
\multirow{3}{*}{Datasets} & \multirow{3}{*}{Backbones} & \multicolumn{2}{c}{Detection} & \multicolumn{2}{c}{Segmentation}
& \multicolumn{3}{c}{Classification} \\
\cmidrule(lr){3-4} \cmidrule(lr){5-6} \cmidrule(lr){7-9}

& & mAP & mAP@50 & mAP & mAP@50 & prec & recall & F1
\\

\midrule
\multirow{9}{*}{BUS-BRA} &
Resnet-50 &
$60.1$ & $84.3$ &
$59.0$ & $83.5$ &
$72.1$ & $70.2$ & $70.9$\\
\arrayrulecolor[rgb]{0.7,0.7,0.7} \cmidrule(lr){2-9} \arrayrulecolor{black}

&

\cellcolor[rgb]{0.9, 0.9, 0.9} {dinov2} &
$\underline{60.2}$ & $\underline{85.8}$ &
$\underline{59.6}$ & $\underline{87.0}$ &
$86.8$ & $\underline{87.2}$ & $87.0$\\
&

\cellcolor[rgb]{0.9, 0.9, 0.9} {USFM} &
$57.7$ & $85.5$ &
$56.8$ & $86.6$ &
$\textbf{90.2}$ & $85.3$ & $\underline{87.2}$\\

&

\cellcolor[rgb]{0.9, 0.9, 0.9} DeblMIM &
$54.5$ & $82.2$ &
$53.0$ & $83.5$ &
$84.8$ & $84.8$ & $84.8$\\
\arrayrulecolor[rgb]{0.7,0.7,0.7}
\cmidrule(lr){2-9}
\arrayrulecolor{black}

&
SAM &
$55.2$ & $77.2$ &
$54.7$ & $78.0$ &
$80.1$ & $78.3$ & $79.1$\\

&
MedSAM &
$58.7$ & $83.4$ & $57.4$ & $84.2$ &
$84.9$ & $82.9$ & $83.8$\\
&
\cellcolor[rgb]{0.9, 0.9, 0.9} {UltraSam} &
$\textbf{60.7}$ & $\textbf{86.0}$ & $\textbf{60.9}$ & $\textbf{87.2}$ &
$\underline{87.6}$ & $\textbf{87.5}$ & $\textbf{87.5}$\\

\midrule

\multirow{9}{*}{MMOTU2D} &

Resnet-50 &
$19.1$ & $27.4$ & $18.9$ & $27.4$ &
$40.6$ & $40.5$ & $38.4$\\
\arrayrulecolor[rgb]{0.7,0.7,0.7} \cmidrule(lr){2-9} \arrayrulecolor{black}

&

\cellcolor[rgb]{0.9, 0.9, 0.9} {dinov2} &
$\underline{22.8}$ & $\textbf{34.4}$ &
$\underline{22.6}$ & $\textbf{34.2}$ &
$64.0$ & $50.8$ & $42.7$\\

&

\cellcolor[rgb]{0.9, 0.9, 0.9} {USFM} &
$14.7$ & $24.0$ &
$14.8$ & $24.6$ &
$\textbf{68.2}$ & $\underline{59.5}$ & $\underline{61.9}$\\

&

\cellcolor[rgb]{0.9, 0.9, 0.9} {DeblMIM} &
$21.9$ & $33.3$ &
$22.5$ & $33.5$ &
$\underline{64.4}$ & $54.0$ & $56.3$\\
\arrayrulecolor[rgb]{0.7,0.7,0.7} \cmidrule(lr){2-9} \arrayrulecolor{black}
&

SAM &
$19.6$ & $28.8$ & $19.3$ & $28.9$ &
$52.3$ & $51.6$ & $51.2$\\

&
MedSAM &
$18.2$ & $27.3$ & $18.2$ & $28.1$ &
$57.9$ & $50.6$ & $52.2$\\

&
\cellcolor[rgb]{0.9, 0.9, 0.9} {UltraSam} &
$\textbf{23.5}$ & $\underline{33.9}$ & $\textbf{23.5}$ & $\textbf{34.2}$ &
$62.6$ & $\textbf{62.4}$ & $\textbf{62.0}$\\

\midrule

\multirow{9}{*}{GIST514-DB} &

Resnet-50 &
$36.2$ & $56.8$ & $37.3$ & $\underline{60.5}$ &
$\underline{83.4}$ & $\underline{82.3}$ & $\underline{81.9}$\\
\arrayrulecolor[rgb]{0.7,0.7,0.7} \cmidrule(lr){2-9} \arrayrulecolor{black}

&

\cellcolor[rgb]{0.9, 0.9, 0.9} {dinov2} &
$36.8$ & $56.5$ &
$37.7$ & $58.1$ &
$67.0$ & $67.0$ & $67.0$\\

&

\cellcolor[rgb]{0.9, 0.9, 0.9} {USFM} &
$26.6$ & $46.6$ &
$26.8$ & $48.2$ &
$74.4$ & $73.4$ & $73.2$\\

&

\cellcolor[rgb]{0.9, 0.9, 0.9} {DeblMIM} &
$22.3$ & $43.9$ &
$21.6$ & $45.7$ &
$66.8$ & $65.8$ & $65.4$\\
\arrayrulecolor[rgb]{0.7,0.7,0.7} \cmidrule(lr){2-9} \arrayrulecolor{black}
&

SAM &
$\textbf{43.5}$ & $\textbf{66.2}$ & $\textbf{44.0}$ & $\textbf{61.5}$ &
$\textbf{86.7}$ & $\textbf{85.1}$ & $\textbf{84.8}$\\

&
MedSAM &
$34.3$ & $55.0$	& $34.8$ & $53.7$ &
$72.4$ & $70.1$ & $69.1$\\

&
\cellcolor[rgb]{0.9, 0.9, 0.9} {UltraSam} &
$\underline{37.8}$ & $\underline{59.0}$ & $\underline{38.4}$ & $58.0$ &
$78.2$ & $74.2$ & $73.5$\\

\bottomrule
\end{tabularx}

\end{table*}

\section{Results}

\subsection{Interactive Segmentation} 
We aim to determine the most effective approach for fine-tuning SAM on the US-43d dataset. To this end, we present the prompt-based segmentation results in Table~\ref{tab:prompt_seg}, using either the instance's center point or bounding box as prompts. We report mean Average Precision (mAP), which evaluates precision across multiple IoU thresholds (0.5 to 0.95), and mAP@50, which measures precision at a fixed IoU threshold of 50\%. Our experiments show that while adapter-based methods such as the Medical SAM Adapter yield competitive results, full end-to-end fine-tuning consistently achieves the best performance. Notably, fine-tuning does not require additional parameters beyond the base SAM architecture, unlike adapter-based approaches. Therefore, despite the higher GPU training cost, we adopt full fine-tuning for UltraSam throughout the rest of the paper. However, adapters may still be a viable alternative when GPU memory is a limiting factor.
When using point prompts, UltraSam substantially outperforms all baselines, achieving mAP scores of 67.5, 57.5, and 55.5 for BUS-BRA, MMOTU2D, and GIST514-DB, respectively.
SAM struggles with US structures due to domain shifts in its training data.
UltraSam* also demonstrates strong zero-shot performance, except on the GIST514-DB dataset. This can be explained by the fact that GIST514-DB contains full radial views, which differ from the rest of the US-43d data, affecting performance. These trends are illustrated qualitatively in Fig.~\ref{quali}, where UltraSam consistently identifies structures that SAM fail to segment.
With box prompts, SAM and MedSAM improve greatly but still fall behind UltraSam and UltraSam*. UltraSam achieves near-perfect mAP@50 scores (99, 100, and 100 for BUS-BRA, MMOTU2D, and GIST514-DB) and strong mAP results (79.1, 79.5, and 73.0, respectively).

\begin{table*}[t!]

\caption{Prompted image classification (Precision (prec), Recall, F1, \%).}\label{tab:prompt_class}
\begin{tabularx}{\textwidth}{@{\extracolsep{\fill}}c c ccc ccc ccc}
        \toprule
        \multirow{3}{*}{Prompt} & \multirow{3}{*}{Backbones} & \multicolumn{3}{c}{BUS-BRA} & \multicolumn{3}{c}{MMOTU2D} & \multicolumn{3}{c}{GIST514-DB} \\
        \cmidrule(lr){3-5} \cmidrule(lr){6-8}
        \cmidrule(lr){9-11}
        & & prec & recall & F1 &
        prec & recall & F1 &
        prec & recall & F1\\

        \midrule

        \multirow{3}{*}{Point}

        & SAM
        & $81.4$ & $81.4$ & $81.6$
        & $51.3$ & $51.3$ & $51.9$ 
        & $\textbf{86.7}$ & $\textbf{89.4}$ & $\textbf{87.0}$\\

        & MedSAM
        & $-$ & $-$ & $-$
        & $-$ & $-$ & $-$
        & $-$ & $-$ & $-$\\

        & UltraSam
        & $\textbf{88.9}$ & $\textbf{89.1}$ & $\textbf{88.7}$
        & $\textbf{62.7}$ & $\textbf{63.2}$ & $\textbf{62.6}$
        & $76.2$ & $77.6$ & $76.2$\\

        \midrule
        \multirow{3}{*}{Box}

        & SAM
        & $\underline{81.6}$ & $\underline{81.2}$ & $\underline{82.1}$
        & $\underline{52.0}$ & $\underline{52.4}$ & $\underline{51.7}$
        & $\textbf{86.7}$ & $\textbf{89.8}$ & $\textbf{87.1}$\\

        & MedSAM
        & $80.1$ & $\underline{81.2}$ & $80.4$
        & $49.3$ & $50.3$ & $49.5$
        & $\underline{78.5}$ & $78.2$ & $\underline{78.9}$\\

        & UltraSam
        & $\textbf{89.4}$ & $\textbf{89.9}$ & $\textbf{88.9}$
        & $\textbf{62.9}$ & $\textbf{63.4}$ & $\textbf{62.6}$
        & $77.8$ & $\underline{78.6}$ & $77.2$\\
    \bottomrule
    \end{tabularx}

\end{table*}

\begin{table*}[t]

\caption{Object detection, instance segmentation (mAP, \%), and image classification results on GIST514-DB using Mask2Former~\cite{mask2former} with a SAM-ViT backbone. We compare models pretrained on US-43d alone versus those augmented with SA-1B to evaluate the effect of natural image data on out-of-distribution ultrasound performance.}\label{tab:ablation_sa1b_seg}

\begin{tabularx}{\textwidth}{@{\extracolsep{\fill}}c cc cc ccc}
\toprule
\multirow{3}{*}{Pretraining} & \multicolumn{2}{c}{Detection} & \multicolumn{2}{c}{Segmentation} & \multicolumn{3}{c}{Classification} \\
\cmidrule(lr){2-3} \cmidrule(lr){4-5} \cmidrule(lr){6-8}

& mAP & mAP@50 & mAP & mAP@50 & prec & recall & F1
\\

\midrule

US-43d (UltraSam) &
$37.8$ & $59.0$ &
$38.4$ & $58.0$ &
$78.2$ & $74.2$ & $73.5$\\

SA-1B (SAM) &
$43.5$ & $66.2$ &
$44.0$ & $61.5$ &
$86.7$ & $85.1$ & $84.8$ \\

US-43d + SA-1B &
$41.0$ & $60.5$ & $41.4$ & $61.4$ &
$83.4$ & $83.1$ & $83.0$\\

\bottomrule
\end{tabularx}
\end{table*}

\begin{figure*}[t]%
\centering
\includegraphics[width=1.\textwidth]{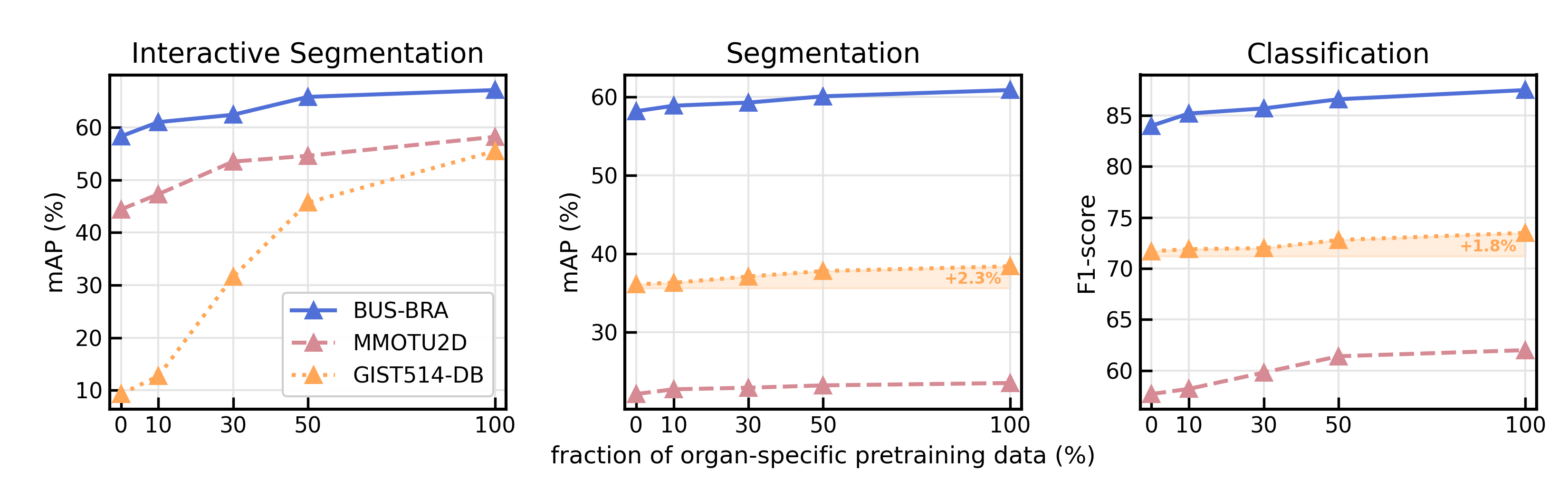}
\caption{UltraSam performance with increasing application-related pretraining data. Performance improves across all test sets as the proportion of related pretraining data increases. Gains are most notable on GIST514-DB, suggesting that adding more radial probe data could further boost performance in this underrepresented domain.}\label{percent_pretraining}
\end{figure*}

\subsection{Downstream tasks}
Table~\ref{tab:downstream_seg} presents results for both instance segmentation (mAP, mAP@50) and image classification (precision, recall, F1-score) across three downstream datasets. On BUS-BRA and MMOTU2D, fine-tuning UltraSam consistently improves over the base SAM, achieving gains of 3–5\% mAP for segmentation and +8.4 and +10.8 F1-score points for classification, respectively. UltraSam also outperforms ultrasound-specific self-supervised models (USFM and DeblMIM~\cite{jiao2024usfm,kang2023deblurring}), confirming the effectiveness of prompt-conditioned pretraining on diverse ultrasound data.

Performance on GIST514-DB follows a different trend. Here, models pretrained on general-domain data, such as SAM, achieve the highest scores for both segmentation and classification. We attribute this to GIST514-DB’s distinctive characteristics, such as its full radial probe views, which are not represented in US-43d. This hypothesis is further explored in Section~\ref{ablation}.

Nonetheless, UltraSam consistently outperforms MedSAM and SSL-based US models across all tasks, including on GIST514-DB, demonstrating its versatility and robustness. These results suggest that prompt-based training on heterogeneous ultrasound datasets provides a strong foundation for downstream transfer, with further gains likely achievable through expanded modality and probe diversity during pretraining.

\subsection{Prompted Image Classification}
The prompted image classification results (Table~\ref{tab:prompt_class}) highlight the benefits of instance-specific prompts, improving upon the base classification task (Table~\ref{tab:downstream_seg}). In the point-prompt setting, the F1 score improves by 1.2 on BUS-BRA, 0.6 on MMOTU2D, and 2.7 on GIST514-DB. Similar improvements are observed for SAM, which remains the best-performing model on GIST514-DB, improving by 2.2.\\

\subsection{Hybrid pretraining}\label{ablation}
UltraSam outperforms ImageNet-initialized ResNet, SAM, and MedSAM on BUS-BRA and MMOTU2D but underperforms SAM on GIST514-DB in downstream tasks. We hypothesize this is due to GIST514-DB’s unique characteristics, such as full radial probe views, which are only present in this dataset within US-43d. Interestingly, ImageNet-pretrained models also outperform MedSAM on GIST514-DB, suggesting these distinct imaging characteristics benefit from broader pretraining.

To investigate, we combined natural images from the SA-1B dataset~\cite{sam} with US-43d in a 50/50 split for pretraining and fine-tuned the model for downstream tasks. This allowed us to compare SAM models pretrained on natural images, ultrasound images, or a mix of both. Results for classification, detection, and instance segmentation (Table~\ref{tab:ablation_sa1b_seg}) on GIST514-DB show that combining natural and ultrasound images in pretraining improves performance over ultrasound-only pretraining. These findings support our hypothesis that GIST514-DB benefits more from natural image pretraining due to its unique characteristics, such as radial probe views, which are not represented in US-43d, and further the gap.

\subsection{Impact of dataset composition}\label{ablation_dataset_composition}
To investigate how dataset composition affects generalization, we conducted an ablation study on the impact of organ-specific data during pretraining. Specifically, we assessed whether including pretraining data related to an application improves performance.
We retrained UltraSam on the full US-43d dataset while varying the proportion of data from organ-specific datasets relevant to each test set: gastrointestinal data for GIST514-DB, breast for BUS-BRA, and ovarian for MMOTU2D. We tested with 0, 10, 30, 50, and 100\% of the available relevant datasets, while keeping all other datasets unchanged. For example, for GIST514-DB, 0\% corresponds to zero-shot setting, and 100\% includes the full available radial GIST training data. 
We present the results for each tasks and test dataset in Fig.~\ref{percent_pretraining}. GIST514-DB exhibits substantial performance gains, particularly in interactive segmentation, as more radial data is introduced during pretraining. Downstream segmentation and classification tasks also improve meaningfully, confirming that foundation model performance benefits from increased exposure to underrepresented probe types. In contrast, BUS-BRA and MMOTU2D show more gradual improvements, likely due to their greater visual similarity to other datasets in US-43d.  
These findings underscore that while prompt-based pretraining enables strong zero-shot generalization, further gains can be achieved through targeted inclusion of underrepresented anatomical regions and acquisition modalities.

\section{Discussion and Conclusion}\label{conclusion}

In this work, we introduced UltraSam, a SAM-style foundation model for ultrasound imaging, trained on our proposed US-43d, the largest compilation of open-access US segmentation datasets to date. We demonstrate that prompt-conditioned segmentation provides a scalable solution for representation learning from highly heterogeneous, sparsely annotated data without requiring dense annotations or unified labeling. UltraSam excels both as an interactive segmentation tool, outperforming existing SAM variants and US-specific models, and as a robust initialization method that significantly enhances downstream tasks such as classification, instance segmentation, and our novel prompted classification task, surpassing SSL-based models initialization.

Nonetheless, we observed performance limitations on datasets with distinct characteristics such as GIST514-DB, due to its unique radial probe imaging that is underrepresented in US-43d. Our ablation study demonstrated that targeted inclusion of organ- or modality-specific data during pretraining has the potential to significantly improve model robustness. Additionally, we showed that hybrid pretraining, combining ultrasound-specific data (US-43d) with natural image data (SA-1B), further mitigates domain-specific knowledge loss, preserving broader visual priors and enhancing performance on such challenging datasets. These findings emphasize the importance of dataset composition and visual diversity in achieving robust US foundation models.

With the release of US-43d, the pretrained UltraSam checkpoint, and our code, we hope to provide valuable resources that advance future research in the field. We encourage the research community to contribute additional high-quality US datasets, especially in underrepresented areas, to improve the model's adaptability across diverse applications.

\section{Declarations}

\noindent{\bf Acknowledgements and Funding}
This research was supported by the ARC Foundation (www.fondation-arc.org) within the APEUS project. This work was also supported by French state funds managed within the ’Plan Investissements d’Avenir’ funded by the ANR under references ANR-21-RHUS-0001 (DELIVER), ANR-20- CHIA-0029-01 (AI4ORSafety) and ANR-10-IAHU-02 (IHU Strasbourg). This work was performed using HPC resources from GENCI–IDRIS (Grant AD011013698R3).

\noindent{\bf Disclosure of potential conflicts of interest}
The authors declare no conflict of interest.

\noindent{\bf Consent to participate}
No informed consent was required as the study did not involve human or animal participants.

\bibliographystyle{sn-basic}
\bibliography{sn-biblio-save}

\end{document}